\documentstyle[preprint,aps]{revtex}
\input epsf
\begin{document}
\tighten
\title{
Probing $\bar u/\bar d$ Asymmetry in the Proton via\\
$W$ and $Z$ Production}

\author{J. C. Peng and D. M. Jansen}
\address{ Physics Division, Los Alamos National Laboratory\\
Los Alamos, New Mexico  87545}

\maketitle
\begin{abstract}
The sensitivity of $W$ and $Z$ production at RHIC to the possible
$\bar u/\bar d$ asymmetry in the proton is studied. The ratios of the
$W^+$ over $W^-$ production cross sections in $p + p$ collision, as well
as the ratios of the $W^+$ and $Z$ production cross sections for $p + p$
over $p + d$ collisions, are shown to be sensitive to this asymmetry.
Predictions of various theoretical models for these ratios are presented.
\end{abstract}

\vfill
\eject

\baselineskip 20pt

The validity of the Gottfried sum rule (GSR) [1], proposed
more than 25 years ago, was
recently tested by the New Muon Collaboration (NMC) in a muon deep inelastic
scattering (DIS) experiment [2].  The Gottfried sum is given as

\begin{eqnarray}
S_G = \int_{0}^{1} dx \, [F_{2}^{\mu p} (x) -F_{2}^{\mu n} (x)] /x = {1 \over
3} +
{2 \over 3} \int_{0}^{1} dx \, {[\bar u(x) - \bar d(x)]}.
\end{eqnarray}

\noindent The second term in Eq. (1) vanishes if
the antiquark distributions in
the proton are SU(2) flavor symmetric, and the Gottfried sum rule, $S_G = 1/3$,
is
obtained.  Based on their measurements of muon DIS on hydrogen and deuterium
targets and an estimate of the contributions from the unmeasured small
$x$-region, the NMC found $S_G = 0.235 \pm 0.026$, significantly different from
the value of 1/3 predicted by the Gottfried sum rule.

A number of theoretical models [3-13] have been proposed to account for the
apparent violation of the GSR. Some models~\cite{c3,c4}
assume that the  valence quark distributions in the proton are sufficiently
singular at small $x$ such
that a large contribution to the Gottfried sum occurs at
a region not probed by the NMC experiment.  Therefore, the GSR is not violated
and the assumption of  $\bar u/\bar d$ symmetry in the proton remains valid.
Other theoretical models [5-13] accept the NMC result as an evidence that the
$\bar u$ and $\bar d$ distributions in the proton are different.  Empirical
expressions  for $\bar d(x) - \bar u(x)$ have been proposed~\cite{c6,c9,c13}
and several recent sets of  parton distribution functions~\cite{c14,c15}
explicitly  allow $\bar u/\bar d$ asymmetry  to accomodate the NMC result.
The origin of the enhancement of $\bar d$ over $\bar u$ in the proton has  been
attributed to pion cloud~\cite{c5,c7,c8,c12}, diquark clustering  in the
nucleon~\cite{c10}, as well as Pauli-blocking effect~\cite{c11}.

Eq. (1) shows that GSR is sensitive only to the integral of $\bar u - \bar
d$.  To provide additional inputs to the possible $\bar u/\bar d$
asymmetry in the proton, it is important to measure the $\bar u/\bar d$ ratio
as a function of $x$.  The proton-induced Drell-Yan process can probe the $\bar
u/\bar d$ asymmetry [6,16,17].  Indeed, the E772
Drell-Yan data have been compared with
predictions from various $\bar u/\bar d$ asymmetric models [18].
More recently, the
NA51 collaboration has measured the ratio of Drell-Yan cross sections on
hydrogen
versus deuterium targets and the result indicates a large asymmetry of $\bar
u/\bar d = 0.51 \pm 0.04 \pm 0.05$ at $x = 0.18$ [19].  Another
Drell-Yan experiment
covering a wider $x$ range $(0.05 < x < 0.3)$ has also been proposed [20].

In addition to the DIS and the Drell-Yan processes, there are other
interactions sensitive to the sea-quark distributions in the nucleon.  One
example is the proton-induced $J/\psi$ and $\Upsilon$ production at a kinematic
region where the cross section is dominated by the quark-antiquark
annihilation [21].  The opportunity to study $p + p$ and $p + A$
collisions at the future
heavy ion collider, RHIC, suggests yet another process, namely the production
of $W$ and $Z$ bosons, which is sensitive to the $\bar u/\bar d$ asymmetry.
In this paper, the sensitivity of $W$ and $Z$ production to the sea quark
distribution in the nucleon is studied.

The differential cross section for $W^+$ production in hadron-hadron collision
can be written as [22]

\begin{eqnarray}
{d \sigma \over dx_F} (W^+) & = & K {\sqrt 2 \pi \over 3} G_F
\left({x_1 x_2 \over {x_1 + x_2}}\right)
\left\{ \cos^2 \theta_c \,
[u(x_1) \bar d(x_2) + \bar d(x_1) u(x_2)] + \right. \nonumber \\
&  & \left. \sin^2 \theta_c \,
[u(x_1) \bar s(x_2) + \bar s(x_1) u(x_2)] \right\},
\end{eqnarray}

\noindent where $u(x), d(x),$ and $s(x)$ signify the up, down, and strange
quark
distribution functions in the hadrons.  $x_1, x_2$ are the fractional momenta
carried by the partons in the colliding hadron pair and $x_F = x_1 - x_2$.
$G_F$ is Fermi coupling constant and $\theta_c$ is the Cabbibo angle.
The factor $K$ takes into account the
contributions from first-order QCD corrections [22]

\begin{eqnarray}
K \simeq 1 + {8\pi \over 9} \alpha_s(Q^2).
\end{eqnarray}

\noindent At the $W$ mass scale, $\alpha_s \simeq 0.1158$ and $K \simeq
1.323$.  This indicates that higher-order QCD processes are relatively
unimportant for $W$ production.  An analogous expression
for $W^-$ production cross
section is given as

\begin{eqnarray}
{d \sigma \over dx_F} (W^-) & = & K {\sqrt 2 \pi \over 3} G_F
\left({x_1 x_2 \over {x_1 + x_2}}\right)
\left\{ \cos^2 \theta_c \,
[\bar u(x_1) d(x_2) +  d(x_1) \bar u(x_2)] + \right. \nonumber \\
&  & \left. \sin^2 \theta_c \,
[\bar u(x_1) s(x_2) + s(x_1) \bar u(x_2)] \right\},
\end{eqnarray}

It is straightforward to show from Eqs. (2) and (4) that, for $p + \bar p$
collisions, the total cross section for $W^+$ production is identical to that
for $W^-$ production.  Figure 1 shows the $W$ production cross sections
measured at the SPS [23] and Tevatron [24] energies.
Calculations using Eq. (2) and
the MRSD-$^\prime$ structure functions reproduce the magnitude and the energy
dependence of W production cross sections well.

An interesting quantity to be considered is the ratio of the differential
cross sections for $W^+$ and $W^-$ production.  If one ignores the much
smaller contribution from the strange quarks, this ratio can be written as

\begin{eqnarray}
R(x_F) \equiv {{{d \sigma \over dx_F} (W^+)} \over
{{d \sigma \over dx_F} (W^-)}} =
{{u(x_1) \bar d(x_2) +  \bar d(x_1) u(x_2)} \over
{\bar u(x_1) d(x_2) +  d(x_1) \bar u(x_2)}}.
\end{eqnarray}

For $p + p$ collision, it is
evident that $R(x_F)$ is symmetric with respect to $x_F = 0$, namely,
$R(x_F) = R(-x_F)$.  For existing $p + \bar p$ collider
experiments at the SPS and
Tevatron, $R(x_F)$ is predominantly sensitive to the valence quark
distributions in the proton.  This can be readily seen by examining $R(x_F)$
at $x_F >> 0$, namely,

\begin{eqnarray}
R(x_F >> 0) (p + \overline{p} \, \, {\rm collision}) =
{{u(x_1) d(x_2) +  \bar d(x_1) \bar u(x_2)} \over
{\bar u(x_1) \bar d(x_2) +  d(x_1) u(x_2)}} \approx
{u(x_1) \over d(x_1)} {d(x_2) \over u(x_2)}.
\end{eqnarray}

\noindent The terms containing $\bar u(x_1)$ and $\bar d(x_1)$ were dropped
since the sea quark distributions are negligible at large $x_1$.
One might also note that $R$ is
equal to 1 at $x_F = 0$. The recent CDF measurement [25] on the lepton
asymmetry
in $W$ production has indeed provided an accurate determination of the
$u/d$ valence quark distributions at $x \approx 0.1$ [26].

The situation is very
different for $p + p$ collision, where $R(x_F)$ becomes sensitive to the sea
quark
distributions in the proton.  Again, it is useful to consider the kinematic
region $x_F >> 0$, where

\begin{eqnarray}
R(x_F >> 0) (p + p \, \, {\rm collision}) =
{{u(x_1) \bar d(x_2) +  \bar d(x_1) u(x_2)} \over
{\bar u(x_1) d(x_2) +  d(x_1) \bar u(x_2)}} \approx
{u(x_1) \over d(x_1)} {\bar d(x_2) \over \bar u(x_2)}.
\end{eqnarray}

\noindent At $x_F = 0$, where $x_1 = x_2 = x$, one obtains

\begin{eqnarray}
R(x_F = 0) (p + p \, \, {\rm collision}) =
{{u(x) \bar d(x) +  \bar d(x) u(x)} \over
{\bar u(x) d(x) +  d(x) \bar u(x)}} =
{u(x) \over d(x)} {\bar d(x) \over \bar u(x)}.
\end{eqnarray}

\noindent As the $u(x)/d(x)$ ratios
are already well determined from the CDF data,
a measurement of $R(x_F)$ in $p + p$ collision gives an accurate determination
of the ratio $\bar d(x)/\bar u(x)$.

Figure 2(a) shows the predictions of $R(x_F)$ for $p + p$ collision at $\sqrt s
=
500~ GeV$.  Five different structure function sets together with the full
expressions for $W^+,W^-$ production cross sections given by Eqs. (2) and (4)
have been used in the calculations.  The solid curve corresponds to the $\bar
u/\bar d$ symmetric DO1.1 structure functions [27], while the dashed and dotted
curves are results for the $\bar u/\bar d$ asymmetric structure function sets
MRSD-$^\prime$ and CTEQ2pM, respectively.  These two structure function sets
were
obtained from recent global fits to Drell-Yan and DIS data including the NMC
result [14,15,28].  Finally, the calculations
using the parameterization  for $\bar
d(x)-\bar u(x)$ given by Ellis and Stirling [6] and Eichten {\it et al.} [13]
are also
shown in Figure 2(a).  The parameterization for $\bar d(x)-\bar u(x)$ were
given
at $Q^2$ of 4 GeV$^2$, and it was evolved to the $W$ mass by using the
Altarelli-Parisi equation for flavor non-singlet structure functions [29].

Figure 2(a) shows that the measurement of $R(x_F)$ for $W$ production in $p +
p$
collision at RHIC could provide a sensitive test of various models.
In contrast,
$R(x_F)$ for $p + \bar p$ collision is insensitive to the $\bar u/\bar d$
asymmetry.  This is illustrated in Figure 2(b), where calculations using
various structure function sets are shown.

Another observable sensitive to the $\bar u/\bar d$ asymmetry is the ratio of
$W^+$ production cross sections for $p+p$ and $p+d$ collisions, which could
also be
studied at RHIC.  The ratio $R^\prime (x_F)$ is defined as

\begin{eqnarray}
R^\prime (x_F) \equiv 2 {{{d \sigma \over dx_F} (p + p \rightarrow W^+)} \over
{{d \sigma \over dx_F} (p + d \rightarrow W^+)}} \approx
{{u(x_1) \bar d(x_2) +  \bar d(x_1) u(x_2)} \over
{u(x_1) [\bar u(x_1) + \bar d(x_2)] +  \bar d(x_1) [u(x_2) + d(x_2)]}}.
\end{eqnarray}

\noindent It is interesting to note that at large $x_F$,
$R'(x_F)$ is probing the $\bar
u/\bar d$ ratio, while $R'(x_F)$ is sensitive to $u/d$ at very negative $x_F$.
More specifically,

\begin{eqnarray}
R^\prime (x_F) \approx 1 + \left({{\bar d(x_2) - \bar u(x_2)} \over
{\bar d(x_2) + \bar u(x_2)}}\right) \, \, {\rm at} \, \, x_F >> 0; \nonumber \\
R^\prime (x_F) \approx 1 + \left({{u(x_2) - d(x_2)} \over
{u(x_2) + d(x_2)}}\right) \, \, {\rm at} \, \, x_F << 0.
\end{eqnarray}

\noindent Figure 3 shows the
predictions of $R'(x_F)$ for various models using the full
expression for $W^+$ production given by Eq. (2).  The divergence of the
various curves at large $x_F$ reflects the different prescriptions of $\bar
u/\bar d$ for the various models.  The predictions converge at negative
$x_F$, showing that the various models have very similar parameterizations for
the well determined valence quark distributions.

Finally, the $Z^\circ$ production could also provide information on the
$\bar u/\bar d$ asymmetry  in the proton. The $Z^\circ$ differential
cross section is given as

\begin{eqnarray}
{d \sigma \over dx_F} (Z^\circ) & = & K {\pi \over 3 \sqrt{2}} G_F
\left({{x_1 x_2} \over {x_1 + x_2}}\right) \left\{
(1-{8\over3}\chi_w + {32\over9} \chi_w^2) \,
[u(x_1) \bar u(x_2) + \bar u(x_1) u(x_2)] + \right. \nonumber \\
& & \left. (1-{4\over3}\chi_w + {8\over9} \chi_w^2) \,
[d(x_1) \bar d(x_2) + \bar d(x_1) d(x_2) +
s(x_1) \bar s(x_2) + \bar s(x_1) s(x_2)] \right\}.
\end{eqnarray}

\noindent $\chi_w$ denotes $\sin^2 \theta_w$, where $\theta_w$ is
the Weinberg angle.
The $K$-factor at the $Z^\circ$ mass scale is equal to 1.317, according to
Eq. (3). Using Eq. (11) and the MRSD-$^\prime$ structure function,
the prediction for $Z^\circ$ production in $p + \overline{p}$ collisions is
shown in Figure 1. The $Z^\circ$ production cross sections measured at UA2
and CDF are well reproduced.

For $p + p$ collision at large $x_F$, the $u(x_1) \bar u(x_2)$ term in
Eq. (11) gives the dominant contribution. The ratio of Z$^\circ$
production cross sections in $p + p$ and $p + d$ collisions can be written as

\begin{eqnarray}
R''(x_F) \equiv 2 {{{d \sigma \over dx_F} (p + p \rightarrow Z^\circ)} \over
{{d \sigma \over dx_F} (p + d \rightarrow Z^\circ)}} \approx
1 - \left( \bar d(x_2) - \bar u(x_2) \over \bar d(x_2) + \bar u(x_2) \right).
\end{eqnarray}

\noindent Figure 4 shows the predictions of $R''(x_F)$ for various models.
For the $\bar u/\bar d$ symmetric model, DO1.1, $R''(x_F)$ approaches one
for $x_F \ge 0.1$. For the $\bar u/\bar d$ asymmetric models, the values
of $R''(x_F)$ at large $x_F$ directly reflect the amount of
$\bar u/\bar d$ asymmetry.

In conclusion, $W$ and $Z$ production at RHIC would offer an
independent means to examine the possible $\bar u/\bar d$ asymmetry
in the proton. Measurements of the cross section ratios of $W^+$ and
$W^-$ production in $p + p$ collisions, as well as the $W^+$ ($Z^\circ$)
cross section ratios in $p + p$ and $p + d$ collisions, would provide a
sensitive
test of various theoretical models on the sea quark distributions of the
proton. While the Drell-Yan and quarkonium production experiments would
provide important information on the $\bar u/\bar d$ asymmetry
at low $Q^2$, the $W$ and $Z$ production experiments offer the unique
opportunity of probing the possible asymmetry at very high $Q^2$.

\begin{figure}
\vskip 7.5 in
\caption{Cross section times the electron decay branching
ratio for W and Z production from UA2 [23] and CDF [24] collaborations.
The curves are predictions using Eqs. (2) and (11) and the
structure function MRSD-$^\prime$ [25].}
\label{wzfig1}
\vskip -8.5 in
\epsffile{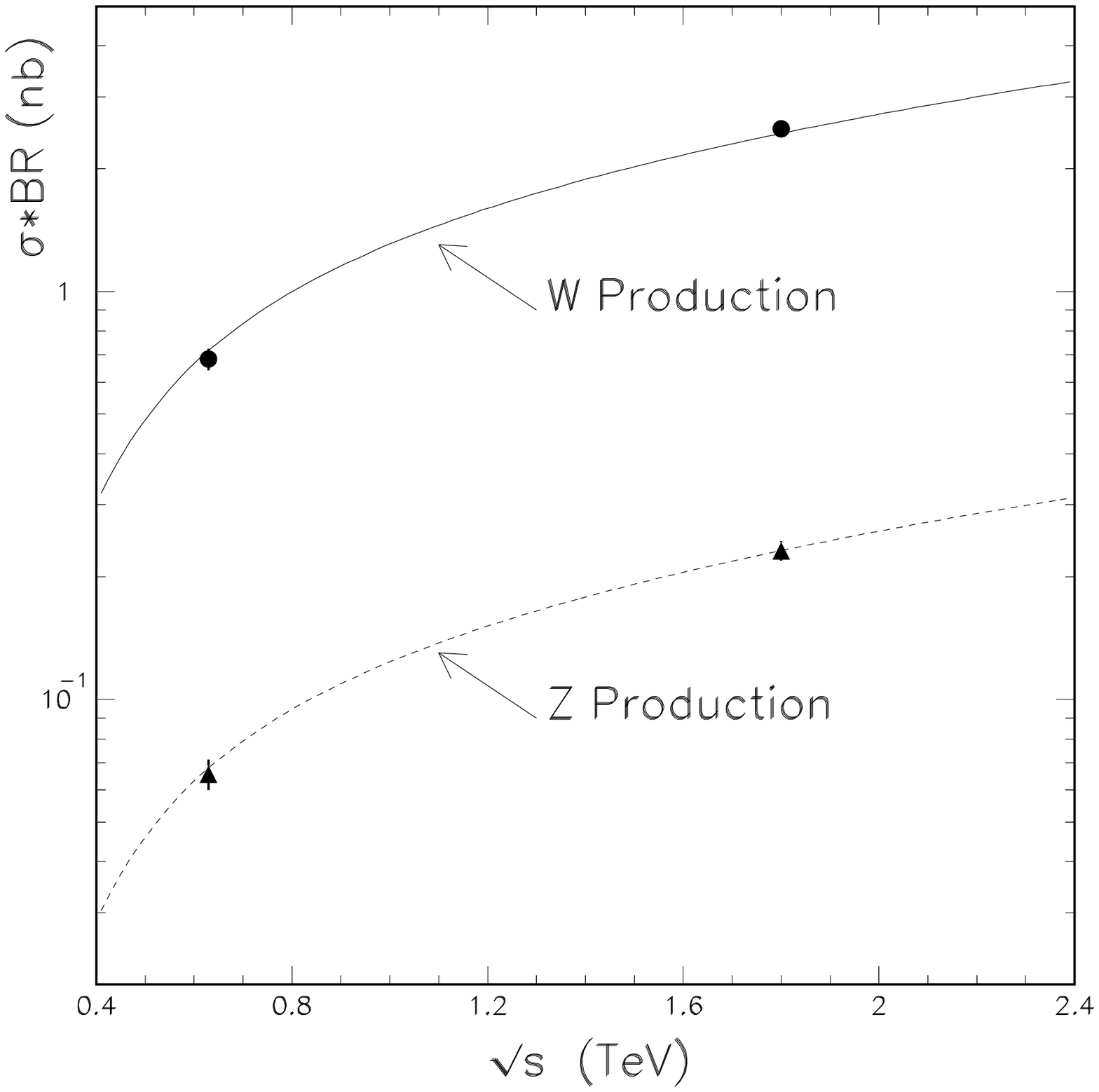}
\end{figure}

\begin{figure}
\epsffile{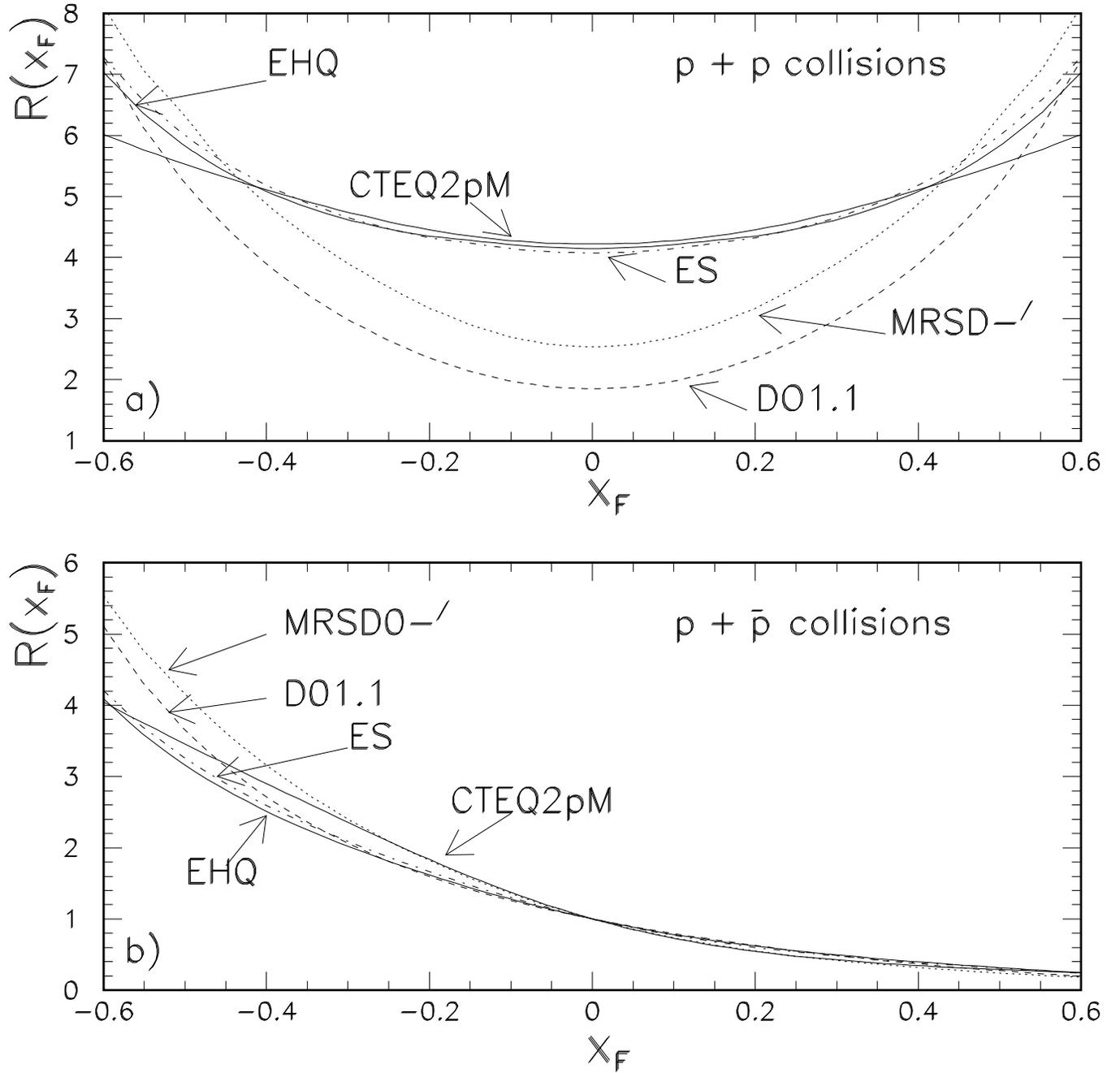}
\caption{Predictions of $R(x_F)$ using Eqs. (2) and (4) and various structure
functions for a) $p + p$ collisions, and b)
$p + \bar p$ collisions at $S^{1/2}$ = 500 GeV.}
\label{wzfig2}
\end{figure}

\begin{figure}
\epsffile{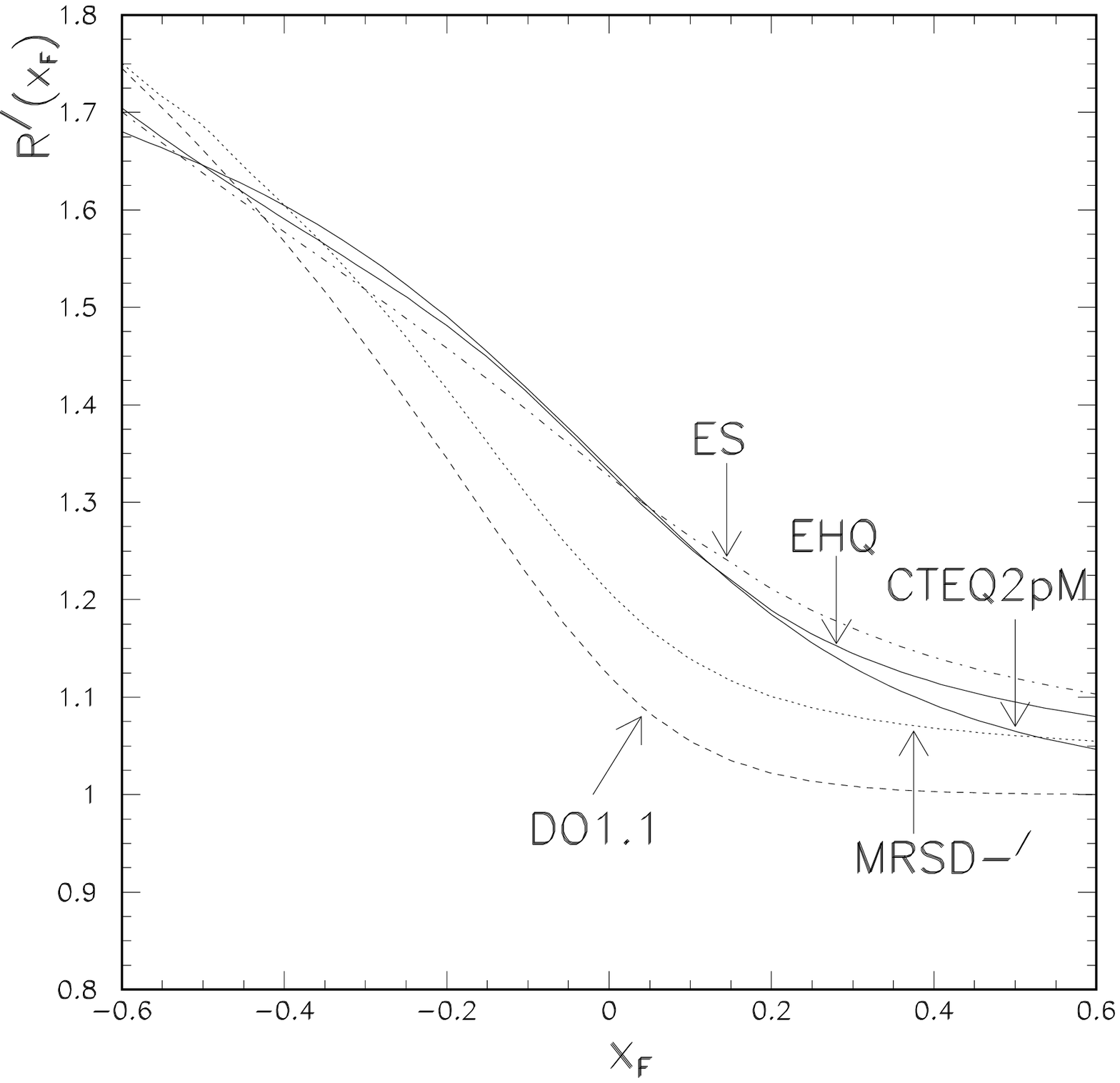}
\caption{Predictions of $R^\prime (x_F)$ using Eq. (2) and various structure
functions for $W^+$ production at $S^{1/2}$ = 500 GeV.}
\label{wzfig3}
\end{figure}

\begin{figure}
\epsffile{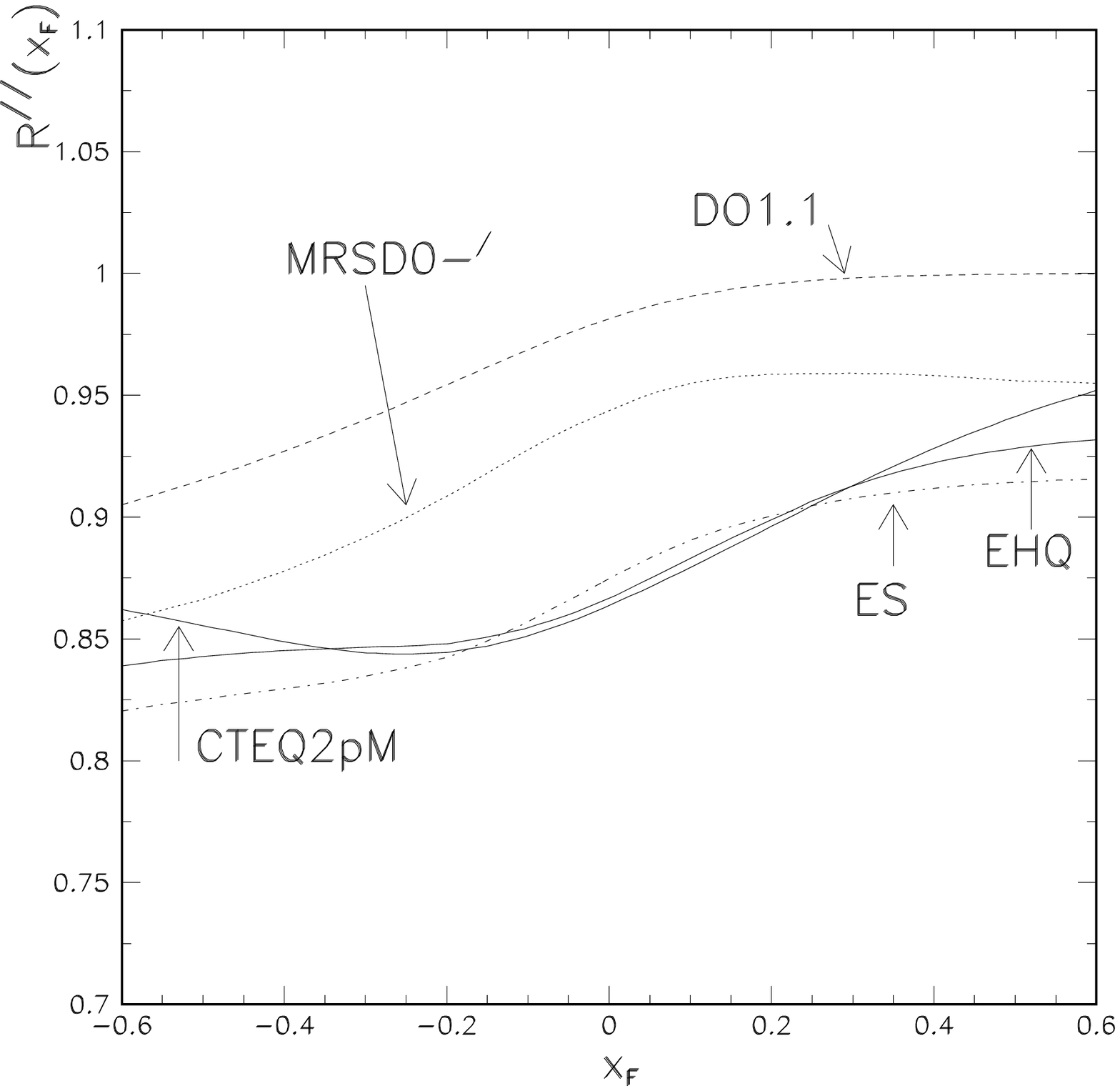}
\caption{Predictions of $R^{''}(x_F)$ using Eq. (11) and various structure
functions for $Z^\circ$ production at $S^{1/2}$ = 500 GeV.}
\label{wzfig4}
\end{figure}

\end{document}